# Room temperature electronic template effect of pre-structured SmSi(111)-8x2 interface yielding self-aligned organic molecules

Younes Makoudi, [a] Eric Duverger, [a] Madjid Arab, [a] Frédéric Chérioux, [a] Franscisco Ample, [b] Gwénaël Rapenne, [b] Xavier Bouju, [b] and Frank Palmino* [a]

*This work describes an innovative concept for the development of organized molecular systems thanks to the template effect of the pre-structured semi-conductive SmSi(111) interface. This substrate was selected because Sm deposition in the submonolayer range leads to a 8x2-reconstruction, which is a well-defined one-dimensional semi-metallic structure. Adsorption of aromatic molecules (1,4-di-(9-ethynyltriptycenyl)-benzene) on SmSi(111)-8x2 and Si(111)-7x7 interfaces has been investigated by scanning tunneling microscopy (STM) at room temperature. Density functional theory (DFT) and semi-empirical (ASED+) calculations have been performed to define the nature of the molecular adsorption sites of the target molecule on SmSi as well as their self-alignment on this interface. Experimental data and theoretical results are in good agreement.*

## Introduction

The fabrication methods of the microelectronics industry have been refined to produce ever smaller devices, but will soon reach their fundamental limits. A promising alternative route to smaller scale functional systems with nanometer dimensions is the autonomous ordering and assembly of atoms and molecules on atomically well-defined surfaces. This approach combines ease of fabrication with fine control over the shape, composition and mesoscale organization of the surface structures formed.[1] Once the mechanisms controlling self-ordering phenomena are fully understood, the self-assembly and growth processes can be accounted to create a wide range of surface nanostructures from metallic, semiconducting and molecular materials. The development of metallic[2] or organic nanolines[3] on a surface is one of the major challenges in nanoelectronics because this type of nano-objects is one of the elemental building blocks for nanocircuits. In order to develop self-directed growth of one-dimensional (1D) molecular lines, two strategies have been investigated.

The first one is based on the supramolecular chemistry which is characterized by a control of the self-assembly process and an intrinsic defect tolerance.[4] The dimensions of the molecular network depend on the geometry of each building block, on the number of noncovalent intermolecular interactions, and on the balance between molecule/molecule and molecule/substrate interactions. As a consequence, the formation at room temperature of 1D molecular lines is expected much more difficult because the number of molecule/molecule interactions is lower than in two-dimensional self-assemblies.[5] Most of the published work has indeed achieved to lessen the molecule/surface interactions which can be prejudicial to the supramolecular self-assemblies on metallic surfaces. The second strategy is based on the guided growth of a molecular line due to a template effect of the surface. In this case, the molecule/surface interactions are predominant. For example, this approach leads to the formation of covalently bonded 1D functional molecular lines along,[6] across[7] the dimer rows of H-terminated Si(001), or on the H-free Si dimers of this surface.[8] However, this second approach is based on the formation of at least one covalent C-Si σ-bond between molecule and substrate, which can induce strong modifications of the electronic structures of the molecules.

In this paper, we show that a molecular alignment can be obtained on a specific silicon interface at room temperature without covalent bonds between molecules and substrate and with negligible molecule/molecule interactions. Our concept is based on the template effect induced by the electronic structure of the substrate which exhibits atomic lines with semi-metallic character. The compromise between the migration barrier height, the molecule/substrate interactions and the intermolecular interactions enables to preserve the electronic structure of the molecules constituting the 1D network. We have used pre-structured silicon reconstruction of the SmSi(111) system.[9] Scanning tunneling microscopy (STM) and density functional theory (DFT) calculations demonstrated that the 8x2-reconstruction was the best candidate for the 1D-organization of organic molecules by template effect of this interface. This remarkable property has been highlighted by large scale STM observations at room temperature of 1D-aligment of organic molecules. The experimental data and theoretical results are in very good agreement.

[a] Y. Makoudi, Dr. E. Duverger, Dr. M. Arab, Dr. F. Chérioux and Dr F. Palmino.
Institut FEMTO-ST, CNRS, Université de Franche-Comté, 32 Avenue de l'Observatoire, F-25044 Besançon cedex, (France).
Fax: (++ 333 8185 3998)
E-mail: frank.palmino@pu-pm.univ-fcomte.fr

[b] Dr. F. Ample, Dr. G. Rapenne, and Dr. X. Bouju
NanoSciences Group, CEMES-CNRS, 29 rue Jeanne Marvig, BP 94347, F-31055 Toulouse cedex 4, (France).

## Results and Discussion

### 1. Experimental Results

1,4-di-(9-ethynyltriptycenyl)benzene (DETB) is built around a 1,4-diethynylbenzene axle with both sides equipped with a triptycene wheel.[10] This molecule was selected because it is an apolar construction that does not possess hydrogen atoms giving rise to intermolecular H-bond (Figure 1). Therefore, no self-assembled supramolecular organization with these molecules is possible because the intermolecular interactions are too weak.

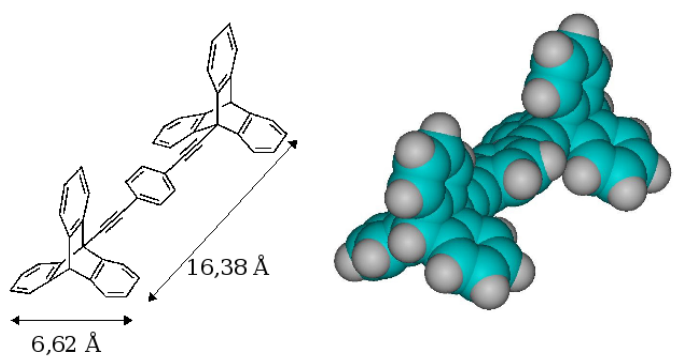

**Figure 1.** Chemical structure (left) and CPK (Corey-Pauling-Koltun) model of the minimum energy conformation (right) of the molecular line subunit DETB.

STM images of this compound have been previously reported on Cu(100) surface but no supramolecular ordering has been observed.[10a] In addition, this molecule was chosen because its geometry is in adequation with the topology of the SmSi-8x2 interface. The total molecular length is 1.64 nm and the distance between two phenyl groups of the triptycenyl moieties is close to 0.66 nm, which corresponds exactly to the distance between two Sm chains of the 8x2-reconstruction perpendicularly to the [11-2] direction.

In order to determine the stability of the molecules on semiconductor, they have been deposited on Si(111)-7x7 that is known to be a highly reactive surface towards organic molecules.[11] A high resolution STM image of isolated molecules adsorbed on this surface for a very low coverage is described in Figure 2.

Molecules appear like two individual protrusions due to the two triptycene wheels. The measured total length of the molecule is 2.2 nm. These results are in a very good agreement with the STM observations performed at low temperature (12 K) with the same molecule on Cu(100)[10a] or a similar compound on Cu(110).[12] The interactions between molecules and Si(111)-7x7 are stronger than molecule/Cu(100) because in the former case STM images are obtained at room temperature, instead of 12 K for the latter case, without destruction of the molecules which is rare on silicon surface.[13] In addition, due to the symmetries of the initial substrate and of the molecules, three sets of possible orientations of the molecules are statistically observed in the same ratio and without alignment on large scale STM image.

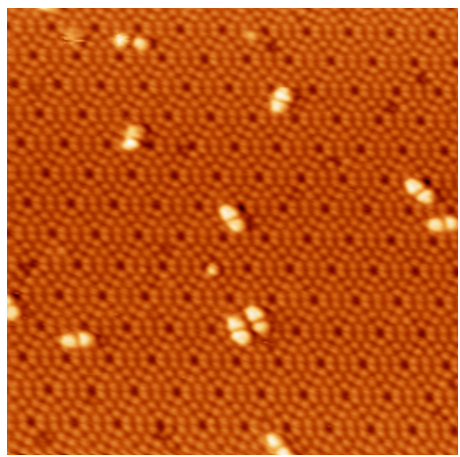

**Figure 2.** High resolution STM image (30x30 nm$^2$) of DETB on Si(111)-7x7, Vs = +2.83 V, It = 0.03 nA.

As observed on Si(111)-7x7 and on Cu(100), molecules on SmSi(111)-8x2 appear like two individual lobes (Figure 3) proving that the molecules are adsorbed on this surface with the same conformation *i.e.* with two couples of phenyl rings directed to the surface as previously described.[10a] In the case of SmSi(111)-8x2, we observe a good self-alignment of molecules along the [1-10] direction and the molecules are parallel to the atomic rows of the 1D SmSi(111)-8x2 structure.

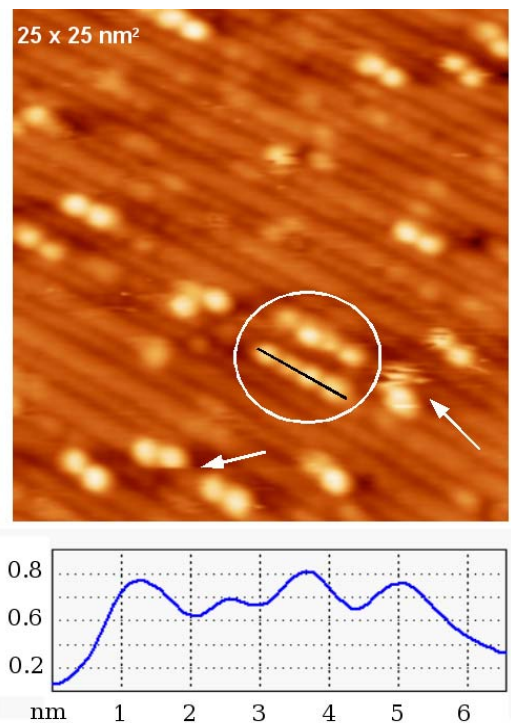

**Figure 3.** STM observation of the DETB subunits on SmSi(111)-8x2 interface at low coverage regime (Vs = +2.3 V, It = 0.02 nA) (top) and profile of the molecular line (down). The Sm lines are also observed on the substrate.

We propose the following scenario to describe the formation of molecular line. After sublimation, the molecules land randomly on the surface and diffuse until they are trapped along a Sm line. Most of molecules are isolated from one another but seeds of molecular lines can already be observed (see white circle, in Figure 3), that contain two consecutive molecules separated by a distance close to 1 nm. Some molecules are not exactly aligned with the Sm atomic rows and a few of them (white arrows, Figure 3) present a characteristic STM molecule shifting during the scanning. These two types of "defects" represent around 10% of the molecules. The room-temperature observation of well-defined molecular contours demonstrates that molecule/surface interactions are stronger on the SmSi interface than on a metallic surface where the individual molecules move continuously at room temperature on atomically flat terraces or can be moved by STM tip.[10a] This important result implies that the molecules have diffused on the substrate at room temperature and that the surface acts as a 1D-template for molecule adsorption. A statistical study shows that 90% of the molecules are perfectly aligned on the large stripes of the SmSi(111)-8x2 structure.

If the coverage rate is increased, the molecules are again adsorbed randomly onto Sm atomic lines of the SmSi(111)-8x2 structure (Figure 4). As found at low coverage regime, it is remarkable that around 90% of the molecules are aligned along the Sm atomic rows. Similarly to the result obtained at low coverage, around 10% of the molecules are not exactly aligned along the Sm atomic rows, probably due to weak intermolecular interactions. This experiment indicates that the major interaction takes place between molecule and substrate. Short molecular nanolines (40% of adsorbed molecules), resulting from molecule/molecule interactions, are observed at high coverage regime. The distance between two parallel lines of adsorbed molecules is 2.6nm (*i. e.* measured between the maxima of the protrusions, Figure 4), which exactly corresponds to the distance between two neighbouring Sm lines. [9b]

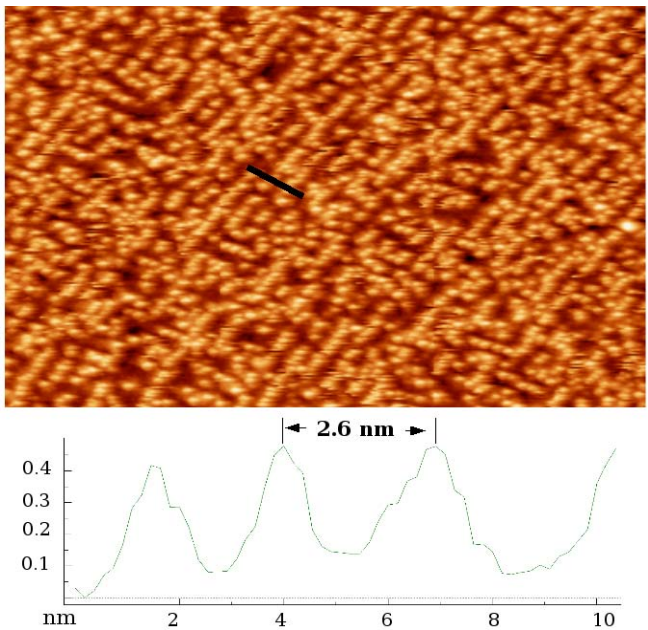

**Figure 4.** Large scale (100x60 nm$^2$) STM image with atomic high resolution of DETB on SmSi(111)-8x2 at high coverage regime (Vs = +2.4 V, It = 0.7 nA) and scan line corresponding to the black line on the STM image.

**2. Theoretical results and discussions**

In order to validate the SmSi(111)-8x2 structural model proposed in previous works[9b] and to find the exact adsorption site of the molecule after deposition, we have performed a complete theoretical study of the substrate by DFT with Vienna *ab initio* simulation package (VASP).[14] The interaction between ions and electrons was described by the projector augmented wave (PAW) method. The generalized gradient corrected approximation (GGA) functional Perdew-Wang 91 and a plane wave cut-off of 250 eV were used. A Monkhorst-Pack k-point grid was used corresponding to roughly 16x16x1 k points in the unit cell. Integrated LDOS (Local Density Of State) were obtained following the Tersoff-Hamann approach,[15] where constant-current images are approximated by an isosurface of the local DOS, integrated between the Fermi energy ($E_f$) of the system and the tip bias voltage ($E_f + V_s$). The integrated LDOS for a sample bias voltage of +2.3 V of the SmSi(111)-8x2 interface is shown in Figure 5 (left).

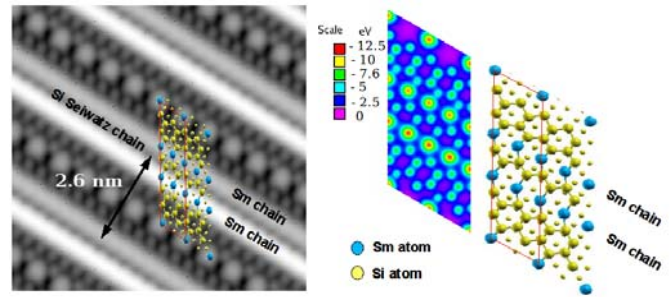

**Figure 5.** Integrated LDOS with a bias voltage of 2.3 V (left) and surface potential of the SmSi(111)-8x2 reconstruction (right).

In integrated LDOS, the Sm rows near the Si Seiwatz chain appear like a large stripe without atomic resolution inside, whereas the x2 Sm atoms periodicity are highly resolved. This result is in good agreement with the empty-states STM observations (see Figure 3 and ref. 9b). In order to understand the poor resolution of the Sm rows, ab initio calculation without spin orbit coupling of the valence and conduction band energy eigenvalues have been performed. In the irreducible Brillouin zone, 10 points have been generated including two high symmetry points Γ, X. (Figure 6)

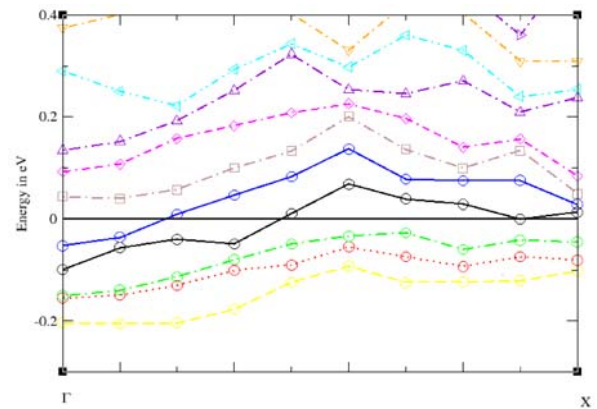

**Figure 6.** Band structure along the principal high-symmetry direction in the Brillouin zone calculated at the predicted equilibrium position. The energy zero is taken at the Fermi level.

Some energy bands go through the Fermi level and we observe a small overlap of the conduction band and valence bands along the Sm rows (*i. e.* [1-10] direction). We suppose that SmSi(111)-8x2 exhibits a semi-metallic character and highlights the poor resolution of the experimental and theoretical STM.[16] Moreover, the surface potential on SmSi(111)-8x2 shows that there is a positive barrier for the Sm atoms situated between two HCC and potential wells onto Sm atoms between HCC and Seiwatz chains. The linear template effect on the Sm lines is evidenced in STM images and it originates from electronic interactions between molecules and Sm atoms rather than surface structure only. Moreover, in the case of a purely structural template effect, the molecules could also be adsorbed onto the hcc structures of Si atoms, which are larger than the molecules. However, this case has never been observed in the experimental STM images, justifying the electronic template effect.

The determination of adsorption properties with DFT codes are particularly time consuming. As the physical system includes 399 atoms, among which 319 atoms for the rigid substrate, the complete system (*i. e.* molecule+substrate) has been simulated by a modified version of the semi-empirical atom superposition and electron delocalization molecular orbital (ASED-MO) approach[17] instead of VASP in order to decrease the time of calculation. Considering the STM images, the molecule has been positioned in order to superimpose two phenyl groups of each triptycene onto Sm chains.

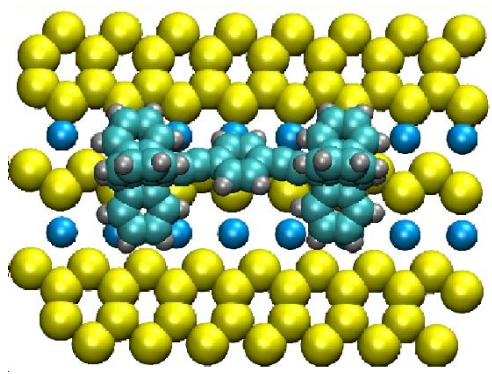

**Figure 7.** Adsorption geometry of DETB on the SmSi(111)-8x2 reconstruction, calculated with ASED+ code.

In Figure 7, deformations of the molecule are clearly identified and are mainly attributed to the near close registry of the molecule along the Sm row. In fact, the distance between the two wheels is reduced by 0.3 Å and the angle between two branches of the wheel closest to the surface is about 130 degrees, compared to the 120 degrees of the molecule without substrate. The axle-surface distance is 3.95 Å and the distance between the extremity of the four wheel blades and the surface is 2.5-2.7 Å according to the strength of the interaction between C and Sm atoms.

The diffusion barrier along the rows has been estimated by the ASED+ code to be around 0.35-0.40 eV per molecule and lateral van der Waals interactions between two neighboring adsorbed molecules in this oriented direction are 10-20 meV. These values are fully consistent with the experimental observation of aligned molecular structures.

## Conclusion

An innovative concept for the 1D self-organization of organic molecules on semiconductor at room temperature has been reported. This concept is based on the template effect of the electronic structure of the substrate which exhibits atomic lines of Sm atoms with semi-metallic character. In the field of nanoelectronics, this new route appears as an elegant strategy to develop supramolecular architectures without intermolecular interactions. Moreover, the connectivity between aligned molecules can be enhanced and work is now underway to functionalize the molecules with end groups[18] able to strongly interact in order to control the construction of molecular wires along Sm lines.

## Experimental Section

STM experiments:

Experiments were carried out in an ultrahigh vacuum chamber with a pressure lower than $10^{-10}$ mbar equipped with a STM (Omicron). Images were recorded at room temperature and acquired in the usual constant current mode. The substrates were n-type Si(111) wafers of 0.1 Ω.cm resistivity. The SmSi interfaces were prepared by Sm evaporation from a Mo crucible onto the Si(111)-7x7 substrate held at 550°C. The evaporation rate was close to 0.63 monolayer (ML) per minute. The monolayer scale is referred to the Si(111) surface atomic density $7.8 \times 10^{14}$ atoms/cm$^2$. The 8x2-reconstruction is obtained for 0.3 monolayer of Sm. The molecules were sublimed at 200°C.

## *Acknowledgements*

*This work was partly funded by the NMP Programme of the European Community Sixth Framework Programme for RTD activities, under the IP project PICOINSIDE, contract No. NMP4-2004-500328, and by the Communauté d'Agglomération du Pays de Montbéliard. Dr Isabelle M. Dixon is warmly acknowledged for her corrections and comments on this manuscript.*

**Keywords:** Surface chemistry · Scanning probe microscopy · Semiconductors · Organic chemistry · Theoretical chemistry

Entry for the Table of Contents (Please choose one layout)

Layout 1:

## ARTICLES

Formation of nanolines of aromatic molecules is achieved at room temperature thanks to electronic template effect of SmSi(111)-8x2 interfaces and investigated by scanning tunneling microscopy. The template effect of the surface is highlighted by DFT calculations, which shows the semi-metallic character along the chains of Sm atoms.

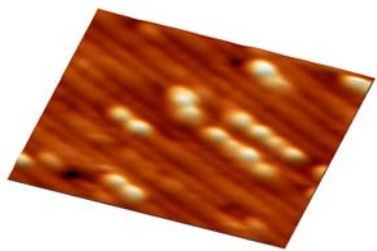

*Y. Makoudi, E. Duverger, M. Arab, F. Chérioux, F. Ample, G. Rapenne, X. Bouju, and F. Palmino\**

*Page No. – Page No.*

**Room temperature electronic template effect of pre-structured SmSi(111)-8x2 interface yielding self-aligned organic molecules**